\documentclass[journal]{IEEEtran}
\usepackage{cite}
\usepackage{xcolor}
\usepackage{float,afterpage,graphicx}
\usepackage{amsmath,amssymb}

\newcommand{\pr}{\text{Pr}}

\ifCLASSINFOpdf
\else
\fi
\begin{document}
%
\title{Bayesian and Algebraic Strategies to Design in Synthetic Biology}
%
%
%

\author{Robyn P. Araujo, Sean T. Vittadello and Michael P.H. Stumpf
\thanks{R.P. Araujo is with the Queensland University of Technology}
\thanks{S. T. Vittadello and Michael P.H. Stumpf are with the University of Melbourne}
}

%
%

\markboth{17. August~2021}%
{Shell \MakeLowercase{\textit{et al.}}: Bare Demo of IEEEtran.cls for IEEE Journals}
%



\maketitle

\begin{abstract}
Innovation in synthetic biology often still depends on large-scale experimental trial-and-error, domain expertise, and ingenuity. The application of rational design engineering methods promise to make this more efficient, faster, cheaper and safer. But this requires mathematical models of cellular systems. And for these models we then have to determine if they can meet our intended target behaviour. Here we develop two complementary approaches that allow us to determine whether a given molecular circuit, represented by a mathematical model, is capable of fulfilling our design objectives. We discuss algebraic methods that are capable of identifying general principles guaranteeing desired behaviour; and we provide an overview over Bayesian design approaches that allow us to choose from a set of models, that model which has the highest probability of fulfilling our design objectives. We discuss their uses in the context of biochemical adaptation, and then consider how robustness can and should affect our design approach. 
\end{abstract}


%
\IEEEpeerreviewmaketitle

\section{Introduction}
%
%
%
%
\IEEEPARstart{T}{rial} and error coupled to ingenuity have arguably been responsible for many important innovations in biotechnology.  Unfortunately, however, these do not scale. Instead recent progress in synthetic biology is generally based on rational design engineering approaches \cite{Lim:2010p24929,SantosMoreno:2020hf}. 
Design in cellular and molecular biology cannot be based directly on the same physical and chemical principles at work in technological systems. Naturally evolved systems are just too complex; our knowledge of their structure and dynamics too primitive; and our ability to interfere with and control their behaviour is still very limited.
\par
Here we discuss a set of complementary approaches that can be used in order to design synthetic biological systems that exhibit certain types of behaviour, or that fulfil certain types of biological functions. We need to acknowledge the differences between biological and technical systems and our ignorance of the details of the molecular mechanisms driving important cellular functions and phenotypes. In light of this we need better ways of developing meaningful, realistic, reliable, and validated mathematical models. And in the context of synthetic biology we need quick and robust ways of picking those models which can meet our design targets and objectives. The latter is the main purpose of this paper.

\par
Below we first outline the problem of design in synthetic biology: we are interested in dynamical systems, implemented in synthetically engineered molecular circuits, that fulfil certain design objectives. One overarching problem is that they need to function robustly and reliably on top of the cellular physiology. We use modelling or {\em in silico} analysis as a way of triaging the vast potential design space in order to identify the best designs, and we will attempt to outline this problem and the confounding mathematical factors in the necessary detail, before illustrating two complementary strategies of exploring this design space. One way of assessing the promise of designs is to assess their level of robustness to uncertainties in the models that we investigate, and  we will make more precise what we mean by robustness. 
\par
We will provide an overview of how algebraic approaches can be used to distill design principles for, at least, certain types of behaviour. In particular, we illustrate how such an approach helps to identify absolute requirements for a dynamical system to exhibit a fundamentally important network response known as Robust Perfect Adaptation (RPA) \cite{Araujo:2018ce}. Algebraic approaches are capable of yielding qualitative insights; Bayesian design methods provide an alternative approach\cite{Barnes:2011hh}. Unlike the algebraic approaches they are computationally costly, but provide rational criteria for choosing from among competing models. We illustrate in general terms how this can be used in practice, again in the context of adaptive responses. 
\par
We conclude by a discussion how robustness of mathematical models can reassure us that a design may function in practice. We argue that taken together, algebraic methods, Bayesian design, and robustness, offer enormous scope for enabling rational design in synthetic biology. 

\section{Design in Synthetic Biology: Objectives and Strategies}
In synthetic biology we are interested in identifying a (dynamical) system that reliably produces a certain type of target behaviour. For example, for a given input, $x$, our system state $y$ should produce a target output or behaviour, $T$. Both $x$ and $T$, can be vector-valued though often they may not be; but $y$, the system state, will be a vector, including all the relevant molecular species (mRNA, proteins, post-translationally modified proteins, metabolites, etc.) involved in a given process. 
\par
We consider systems of the following form,
\begin{equation}
    \frac{d y}{dt} = f(y;x,t,\theta)
\end{equation}
where $\theta$ denotes the system parameters (which are typically unknown). We want to determine if the system given by $f(y;x,t,\theta)$ is capable of producing the desired output, $T$, and meet the design objectives. 
\par
 In our applications we may be interested in systems that can exhibit switch-like behaviour, filter out high frequency noise, exhibit perfect adaptation, or more complex phenotypes, such as the ability of a cellular system to produce Turing pattern behaviour. Specifying the design objectives, $T$, can involve qualitative characteristics, or quantitative aspects of system behaviour. Often we may resort to surrogate data that captures the desired behaviour; this is especially the case if precise mathematical or algebraic definitions of system behaviour are  difficult; or if there are no simple qualitative features (such as algebraic identities).
\par
Model-based design in synthetic biology can be done in different ways: (i) we can use domain expertise to formulate mathematical models, $f(y;x,t,\theta)$, determine if they fulfil our design objectives, and optimise the performance by varying $\theta$ or by making small alterations to the model; (ii) choose from a set of models $\mathcal{F}=\{f_1,f_2,\ldots,f_N\}$ using, for example statistical model selection in order to determine which models are best able to deliver the design objectives;  (iii) explore exhaustively a class of relevant models. That is, now $\mathcal{F}$ might contain all models, $f$, with 2, 3, or 4 species or nodes (the dimension of the vector $y$) to determine if they meet the design objectives; (iv) identify universal characteristics guaranteeing the design objectives are met and construct models (and, as the end-target, synthetic biological systems) embodying these principles. The first approach relies a lot on luck in getting the initial model right. The second and third approach can become  computationally prohibitively expensive even for small networks; they are, however, flexible and widely applicable. The final approach is less {\em ad hoc} than the others, and can provide guarantees on the behaviour of designs. But it requires a more grounded theoretical understanding of the design objectives than we often have.
\par
Of course, there are also ways in which we can combine the last three approaches to gain confidence in our analyses and designs, and to streamline the design process. If, for example, design principles are known we can use any such theoretical principles to reduce the set of potential candidate models, $\mathcal{F}$, potentially quite considerably. 

\section{Algebraic Approaches to Discovering Design Principles}
In this section we consider algebraic approaches that may be able to identify, within the vast potential design space, the exceedingly limited regions that contain the `best' designs -- those that can perform an essential biological function in a robust and reliable manner.   We emphasize that such analytical methods may not be applicable to the study of all possible design objectives, and may not be well suited to the elucidation of complex biological behaviours such as multistationarity \cite{Harrington:2013wv} or Turing patterning\cite{Scholes:2019fq}.  But algebraic (and ultimately topological) methods are beginning to show tremendous promise in the context of robust asymptotic tracking problems, and have recently delineated the full solution space for network topologies capable of Robust Perfect Adaptation (RPA) \cite{Araujo:2018ce}.  RPA is is of fundamental significance in biological systems, both natural and synthetic, and has been widely observed in cellular signalling networks \cite{muzzey, Araujo:2018ce, Khammash2021}, sensory systems \cite{Kaupp, Yau}, in the regulation of bacterial chemotaxis \cite{Barkai1997, Alon1999}, bacterial stress response \cite{Toni:2011jya,Joly:2010p25883}, intracellular osmolarity \cite{you}, sigma-70 activity \cite{briatantithetic}, and homeostatic control of plasma mineral concentrations \cite{mairet, ni}, and is thought to play a role in robust patterning during organism development \cite{eldar, benzvi}.  

\subsection{The Control Theory viewpoint on RPA}
RPA is the ability of a system to generate an output that returns to a fixed reference level (its `setpoint') following a persistent change in the system input, with no need for parameter fine-tuning \cite{Araujo:2018ce}.  RPA is thus equivalent to the well-established control theoretic notion of {\it robust steady-state tracking} or {\it constant disturbance rejection}, while absolute concentration robustness (ACR, a special type of RPA \cite{XiaoDoyle}) is equivalent to the control theoretic concept of {\it robustness to disturbances in initial conditions} \cite{Cappelletti}.  This fundamental problem of robust asymptotic tracking of a desired trajectory, while rejecting unwanted disturbances, is one of the defining problems of classical automatic control \cite{byrnesisidori}.  In the 1970s, the landmark studies of Francis and Wonham \cite{FW1, FW2} investigated the necessary controller structures to achieve robust regulation of a system, with internal stability, and established what is now known as the Internal Model Principle (IMP).  By this principle, a control system is able to reject exogenous stimuli or disturbances by incorporating within itself a model of the dynamic structure of the network stimuli.  For RPA, then, the internal model must produce constant signals and is equivalent to the requirement for integral (feedback) control \cite{sontag}.  Thus, in all RPA-capable biochemical reaction networks, we expect an integrator - an internal model of a system that produces constant inputs or disturbances - to be embedded or concealed within the biochemical reaction rates that govern the dynamical response of the network.   

Suppose we have a $n$-dimensional system of interacting molecules, with state vector $x(t)$ representing the concentrations of the molecules.  In general, we seek some change of coordinates that identifies a subsystem, $z$, of the form
\begin{equation}\label{constrained}
\frac{dz}{dt} = p(x)(\hat{x} - c),
\end{equation}
where $\hat{x}$ represents the concentration of the RPA-capable molecule, and $c$ is some function of system parameters.  The function $p$ may be zero-order in its arguments, in which case RPA is `unconstrained' by any other system properties. In all other cases, Equation \ref{constrained} encodes a concept known as {\it constrained integral control}.

\begin{figure}[ht!]
\caption{Two simple motifs from which an internal model may be identified through a simple change of coordinates.  (A) An antithetic integral control motif \cite{briatantithetic}, as discussed by \cite{ferrell2016perfect}.  (B) An incoherent feedforward motif \cite{sontag}.}
\includegraphics[scale = 0.53]{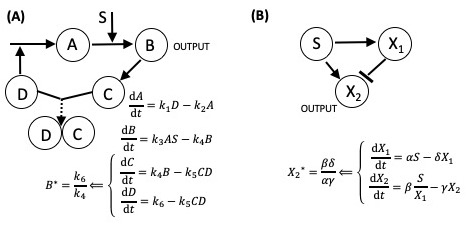}
\label{motifs}
\end{figure}

In simple biochemical network designs, a linear change of coordinates may be sufficient to identify the internal model, such that an integrator of the form in Equation \ref{constrained} lies in the rowspan of the system.  Antithetic integral control \cite{briatantithetic} represents a well-known example of this type and is, topologically speaking, a feedback-regulated system.  A simple representation of this control motif is depicted in Figure \ref{motifs}{A}, with four interacting molecules with concentrations $A, B, C$ and $D$.  From the reaction forms shown, is it clear that 
\begin{equation*}
\frac{dC}{dt} - \frac{dD}{dt} = k_4\left(B - \frac{k_6}{k_4} \right),
\end{equation*}
which captures the internal model explicitly, since  $B^* = k_6/k_4$ represents the setpoint of the molecule $B$.  New methods have recently been developed to detect an internal model in the rowspan of an RPA (or ACR)-capable system, if such exists, in more complex networks of biochemical interactions \cite{Cappelletti, Karp}.

Feedforward structured networks can also be recast as integral feedback systems through a judicious (generally non-linear) change of coordinates.  Such a nonlinear mapping may be difficult to identify for complex feedforward-structured networks, but may be readily obtained for particularly simple feedforward motifs.  For instance, Shoval et al. \cite{shoval} consider a minimal incoherent feedforward motif of the form
\begin{align*}
\frac{dx_1}{dt} &= \alpha u - \delta x_1, \\
\frac{dx_2}{dt} &= \beta \frac{u}{x_1} - \gamma x_2,
\end{align*}
as depicted in Figure \ref{motifs}(B), and show that the map 
\begin{equation*}
\left(X_1, X_2\right) \to \left(z_1, z_2\right) = (X_2, \alpha X_2 - \beta \textrm{log} X_1)
\end{equation*}
 is a diffeomorphism with inverse $X_1 = e^{(\alpha z_1 - z_2)/\beta}$ and $X_2 = z_1$, which transforms the reaction equations into the partitioned form
\begin{align*}
\frac{dz_1}{dt} &= \beta u e^{(z_2 - \alpha z_1)/\beta} - \gamma z_1,\\
\frac{dz_2}{dt} &= \beta \delta - \alpha \gamma z_1.
\end{align*}
The second of these equations  captures the internal model, since the solutions to $dz_2/dt = 0$ represent all possible constant signals, and determine the setpoint of the system to be $z_1 = X_2 = \delta\beta/{\alpha\gamma}$.

The recognition that RPA may be achieved even for very complex networks by adding an integrator to the network within a feedback structure has been of tremendous utility for synthetic implementations \cite{Khammash2021}.   On the other hand, this recognition alone does not provide a straightforward tool for exploring a more comprehensive design space of RPA-capable networks - particularly in view of the fact that the connection to integral feedback regulation may be very complex for networks without a clear (physical) feedback structure (such as feedforward structured networks) \cite{sontag}.  Moreover, there is now strong experimental evidence that RPA in natural (endogenous) cellular chemical reaction networks frequently requires multiple adaptation mechanisms working together in concert \cite{Hoeller}.  Even in the relatively simple chemotaxis-regulating cascades in organisms such as Dictyostelium, for example, it appears that multiple distinct RPA modules are at work \cite{Hoeller}, making it difficult to propose synthetic implementations of these kinds of network structures relying only on the concept of integral control. 

 In the next section, we consider a complementary approach to the integral control viewpoint, by considering how an algebraic representation of the RPA problem may be analysed in such a way as to reveal the complete, comprehensive design space of RPA-capable network topologies.  These methods identify RPA-capable network structures directly, and retain a representation of the RPA problem in terms of the original coordinates (ie. the concentrations of the interacting molecules).  Rather than extract a single internal model for the constant input stimuli to an RPA-capable system, this approach reveals how such an internal model could be {\it distributed} topologically over potentially vast and intricate networks of interacting molecules.

\subsection{The topological RPA problem}
Many algebraic formulations for the RPA problem have been proposed \cite{Araujo:2018ce, Yi, TangMcMillen}, and have been shown to be equivalent to the requirement for integral control \cite{Araujo:2018ce,Yi}.
 Commonly, these formulations allow the network to have multiple `input nodes', ie. molecules that are subjected to stimulation or perturbation by an external source, as well as multiple `output nodes', which are the molecular endpoints of interest, which are expected to exhibit the RPA property.  But to delineate the full solution space of RPA-capable network topologies from such an algebraic condition, it is expedient to start from a simpler assumption:  a network of interacting nodes that are perturbed at a {\it single} input node, and for which a {\it single} output node is monitored for the RPA property, thereby allowing network topologies to be {\it induced} relative to the designated input-output node pair \cite{Araujo:2018ce}.   For an RPA-capable network topology identified by this method, there may exist additional nodes to which a stimulus may be applied while preserving the RPA property at the designated output node.  Likewise, an RPA-capable network design may also exhibit RPA at nodes other than the designated output node.  In this way, a complete picture of RPA-capable multi-input/multi-output network designs emerges. 

The topological RPA problem has now been solved in complete generality \cite{Araujo:2018ce}, and is the fulfilment of a longstanding conjecture in biological network theory \cite{Kitano2004, Kitano2007, AraujoNRDD, araujo2006control, tysonsniffers} which held that robust biosystems are likely to require very specific network architectures - a conjecture based largely on analogies with engineering control systems and the well-defined design features that are known to be required, or at least beneficial, in those contexts.  
The essential ideas underlying this general solution are as follows:

Suppose we have a network consisting of $n$ interacting molecules, or {\it nodes}, denoted $N_1, N_2, \ldots, N_n$.  One of these nodes ($N_i$, say) will be selected as the designated input node, and one node ($N_j$, not necessarily distinct from $N_i$) will be selected as the output.  To each non-input node, we associate a reaction rate $dN_k/dt = f_k(N_1, \ldots, N_n)$, which encodes both the strength and the nature (stimulating or inhibitory) of the inter-molecular interactions upon the node $N_k$.  The input node is also regulated by an external stimulus, $S$, and has a reaction rate $dN_i/dt = f_i(N_1, \ldots, N_n, S)$.  From this simple mathematical representation of the interactions in an $n-$node network, and its encounter with an external stimulus, it is straightforward to show that the network is capable of RPA at node $N_j$ only if
\begin{equation}
\frac{\textrm{Det}(M_{ij})}{\textrm{Det} (J_n)} = 0,
\end{equation}
where $J_n = \frac{\partial(f_1, \ldots, f_n)}{\partial(N_1, \ldots, N_n)}$ denotes the system Jacobian, and $M_{ij}$ is the Jacobian minor obtained by eliminating the input row and the output column from $J_n$.  Thus, a necessary condition for RPA in {\it any} network, regardless of size or complexity, is
\begin{equation}\label{RPA}
\textrm{Det}(M_{ij}) = 0,
\end{equation}
with
\begin{equation}\label{cons}
\textrm{Det} (J_n) \neq 0.
\end{equation}

Equation \ref{RPA} is known as the RPA equation \cite{Araujo:2018ce}, which must be satisfied in any RPA network, subject to the RPA constraint (Equation \ref{cons}).

\begin{figure}[ht!]
\caption{The RPA Equation (Equation \ref{RPA}) grows factorially with the number of interating nodes, $n$ in the network.}
\includegraphics[scale = 0.6]{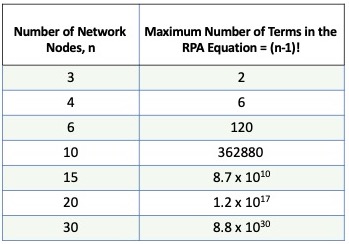}
\centering
\label{nfactorial}
\end{figure}

We note that the RPA equation will, in general, contain a very large number of terms, depending on the number of nodes and the interconnectedness of the network, highlighting the fact that the topological RPA problem quickly eludes computational screening approaches as $n$ grows.  Indeed, for a fully-connected network, where every node regulates every other node, there will be $(n-1)!$ terms in the RPA equation, each containing $(n-1)$ factors.  We illustrate in Figure \ref{nfactorial} how quickly the size of the RPA equation grows with the number of interacting nodes in the network.

\begin{figure*}[ht!]
\caption{Relationship of the factors within individual terms of the RPA Equation to the topology of the underlying network.}
\centering
\includegraphics[width= 0.65\textwidth]{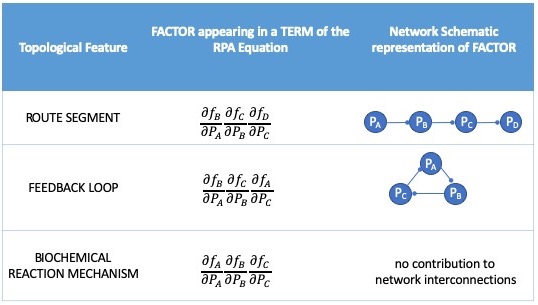}
\label{rpacontent}
\end{figure*}

The cornerstone of the methodology developed in \cite{Araujo:2018ce} is the interpretation of the determinant of $M_{ij}$ not simply as a function Det: $\mathbb{R}^{(n-1)\textrm{x} (n-1)} \to \mathbb{R}$, but as a {\it set} of $\tau$ terms, $1 \leq \tau \leq (n-1)!$, with each term carrying either a positive or a negative sign.  Crucially, the $(n-1)$ factors of each term correspond in a very precise way to topological characteristics of the underlying network, representing either structural features of the network, or constraints on reaction kinetics (see Figure \ref{rpacontent}).

\begin{figure}[ht!]
\caption{Two module classes represent a topological basis for all RPA-capable networks.  RPA-capable nodes are indicated by a red asterisk;  all other nodes are unable to exhibit RPA.  $S_N$ indicates that an arbitrary subnetwork may be embedded at the location indicated without affecting the RPA capacity of the network's output node. A complete description of RPA basis modules is given in \cite{Araujo:2018ce}. (A) An Opposer Module, which involves a feedback structure.  At least one opposer node, which satisfies a special constraint known as {\it opposer kinetics}, must be embedded into the feedback portion of an Opposer Module.  In the representation depicted here, two opposer nodes (indicated in yellow) collaborate to form a special architecture known as a two-node opposing set.  (B) A Balancer Module is characterised by a collection of parallel pathways, with a diverter node (D) at the source point of the parallel pathways, and a connector node (C) at the downstream convergence point. All nodes between D and C, known as balancer nodes (indicated in blue), must satisfy a special constraint known as {\it balancer kinetics}.  The connector node must also satisfy a special constraint known as {\it connector kinetics}. }
\includegraphics[scale = 0.55]{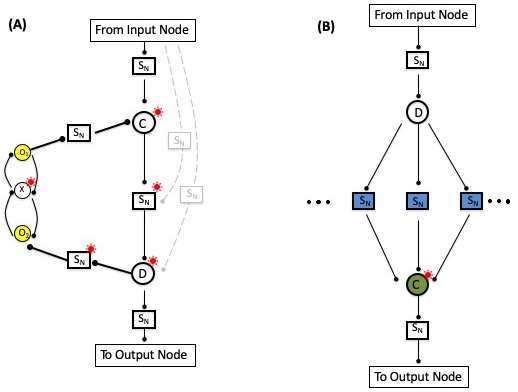}
\label{generalsol}
\end{figure}

It is clear from this set theoretic interpretation of the RPA Equation that partitions of the set may exist for which the contents of every subset can sum to zero independently of the contents of every other subset \cite{Araujo:2018ce}.  By identifying general conditions under which such `independently adapting subsets' can exist, we can determine, definitively, the topological structure that {\it any} RPA-capable network must possess, no matter how large or complex the network.  In particular, it has been shown \cite{Araujo:2018ce} that there exist {\it two and only two} possible types of `independently adapting subset' of the RPA equation, known as S-sets and M-sets.  S-sets contain terms that are all identically zero, for all possible values of input stimulus, and all possible parameter values, and arise when a feedback structure exists within the associated network, and where one or more nodes embedded into that feedback structure satisfy a special constraint on reaction kinetics known as {\it Opposer kinetics}.   These S-sets correspond to the presence of a special subnetwork structure known as an Opposer Module (see Figure \ref{generalsol}(A)).  By contrast, M-sets contain terms that are all strictly non-zero, but together sum to zero for all possible values of input stimulus, and all possible parameter values.  M-sets arise when a collection of parallel pathways exists in the network, and where the nodes embedded into those parallel pathways satisfy a special constraint on reaction kinetics known as {\it Balancer kinetics}.  In addition, the convergence point for these parallel pathways must also satisfy a constraint on reaction kinetics known as {\it Connector kinetics} in order to complete the `balancing act' of the M-set.  M-sets correspond to the presence of a special subnetwork structure known as a Balancer Module (see Figure \ref{generalsol}(B)).  

Opposer and Balancer Modules are generalizations of the simple three-node RPA-capable motifs identified by Ma et al. \cite{Ma:2009p20838} through computational screening methods.  But these more general RPA-promoting modules may be vastly more complex, and can contain previously unknown topological features such as {\it Opposing sets}.  We refer interested readers to \cite{Araujo:2018ce} and \cite{araujoliottamethods} for a comprehensive description of Opposer and Balancer Modules.

These two RPA Modules thus represent a {\it topological basis} for all RPA-capable networks.  In other words, RPA-capable networks are necessarily {\it modular}, and must be decomposable into these well-defined classes of network substructures - an intriguing feature of immense practical utility in the context of synthetic network design.

\begin{figure}[ht!]
\caption{A simple network architecture, which supports two different decompositions into RPA basis modules.  Opposer nodes are indicated in yellow; balancer nodes are indicated in blue; and connector nodes are indicated in green.  It can readily be shown via diagrammatic methods \cite{araujoliottamethods} that the RPA Equation for this network contains three terms, of which two are positive and one is negative.  These terms admit two different partitions into `independently adapting subsets', as indicated schematically beneath each network.  (A)  An Opposer Module emerges as the three terms of the RPA equation are assigned to a single S-set (yellow).  (B) An Opposer Module in series with a Balancer Module emerges as the negative term and one of the positive terms are assigned to an M-set (blue), while the remaining term is assigned to an S-set (yellow).}
\includegraphics[scale = 0.55]{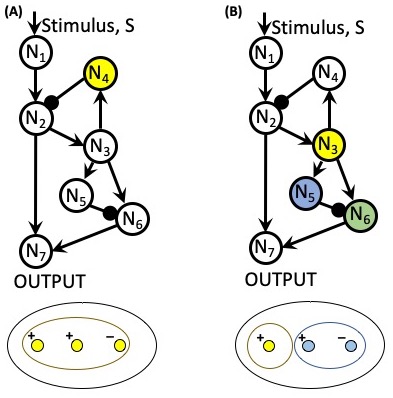}
\centering
\label{singletopology}
\end{figure}

Diagrammatic methods have recently been developed to assist in the analysis of RPA capacity for specific network architectures \cite{araujoliottamethods}.  For the network structure given in Figure \ref{singletopology}, for instance, it can easily be shown that the RPA equation for this particular network contains just three terms, and admits two (and only two) decompositions into RPA basis modules:  a single Opposer Module (Figure \ref{singletopology}(A)), or a Balancer Module in parallel with a (smaller) Opposer Module (Figure \ref{singletopology}(B)).  The rules of interconnectivity for RPA basis modules to create larger, multi-modular networks, and the concept of series and parallel connections of basis modules, are also known in complete generality, and are explained in full in \cite{Araujo:2018ce} and \cite{araujoliottamethods}.

\begin{figure*}[htpb]
\caption{RPA network topologies compute a collection of invariants which, together, confer the RPA property on the output node.  (A) A minimal Balancer Module, in which balancer kinetics at node B compute a {\it balancer invariant}, which is then transferred to node C, which then computes the setpoint.  (B)  A three-node opposing set.  Each of the three independent opposer nodes (yellow) calculates its own {\it opposer invariant}, which is transferred from $O_1$ to $O_2$ to $O_3$, thereby conferring RPA on $P_2$.}
\centering
\includegraphics[scale = 0.6]{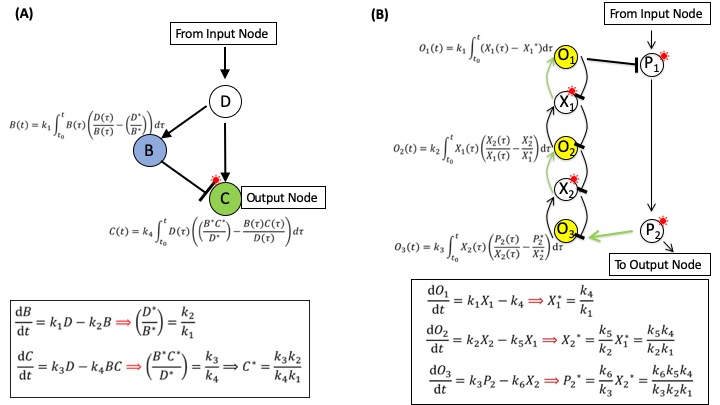}
\label{distributedintegrals}
\end{figure*}

As we close this section, we note that the topological view of the RPA problem brings into sharp relief the concept that biochemical signalling networks do not simply transmit biochemical signals, but are required to perform precise and coordinated {\it computations} on those signals in order to implement robust function.  Indeed, unlike the well-established control theory perspective, which emphasizes the requirement for an internal model (and integral feedback in the case of RPA), topological analysis of robust network designs highlights the presence of a {\it constellation} of invariants, which includes the key invariant referred to as the system setpoint, which are distributed in very specific topological arrangements across a potentially large and intricate signalling network.  Figure \ref{distributedintegrals} illustrates this key idea via a schematic representation of invariant distributions in two minimal RPA modules - a three-node Balancer Module (Figure \ref{distributedintegrals}(A)), and an Opposer Module containing a topological feature known as a three-node opposing set (Figure \ref{distributedintegrals}(B)).  Each node that is subjected to a constraint on its reaction kinetics - a balancer node, an opposer node or a connector node - thereby computes an RPA-promoting invariant in order to confer the RPA property on the output node of the module, and ultimately the entire network.  

\section{Bayesian Perspectives on Design}
Bayesian inference procedures \cite{Robert:2007uy} arguably provides a very natural framework for expressing and reporting uncertainty. Because of this, and its ability to accomodate background information and knowledge it has become an increasingly popular in systems biology \cite{Kirk:2015gj}. One of the main advantages is that it provides a powerful and internally consistent framework for model selection -- that is, choosing from a set of candidate models, those which have a high probability or producing the observed data \cite{Kirk:2013hq}. This can also be used in the context of design for synthetic biology.

\afterpage{
\begin{figure*}
    \centering
    \includegraphics{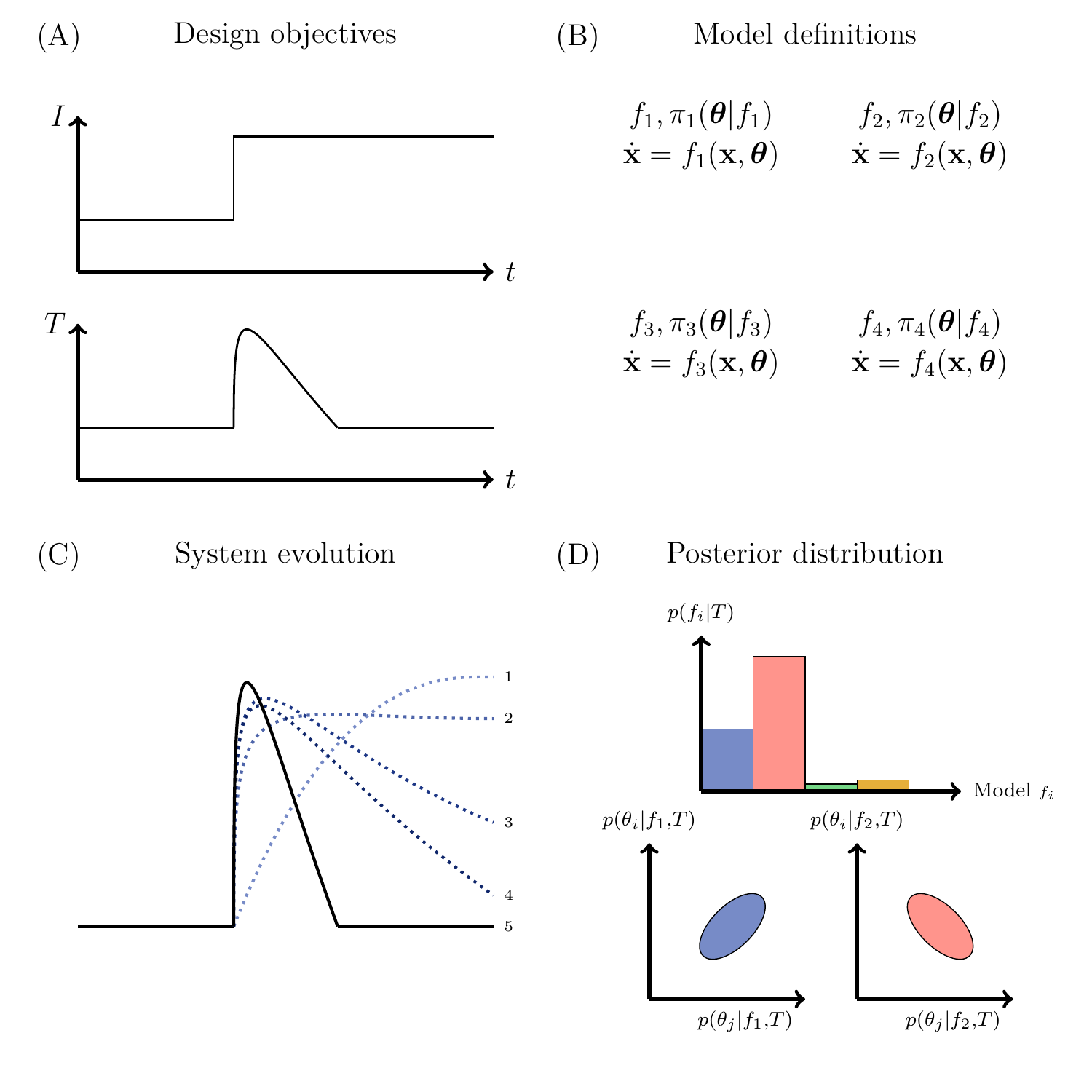}
    \caption{Illustration of the Bayesian Design Framework for Synthetic Biology. (A) The input and desired system output, $T$ for a system exhibiting adaptation. (B) Four alternative models, $f_i(x,\theta)$, are considered, each with a parameter prior, $\pi_i(\theta)$. (C) The ABC-SMC approach begins by considering responses that that increasingly reflect the target objetive, $T$. (D) The output of the ABC-SMC procedure are model posterior probabilities, and for each model parameter posteriors. In the example only the first two models, $f_1$ and $f_2$ have sufficiently high posterior model probabilities and need to be conisdered further.}
    \label{fig:fig_method}
\end{figure*}
}
\subsection{Bayesian Inference for Design}
Bayesian design approaches take a very different, but we feel, complementary approach. Here we consider sets of models $\mathcal{F} = \{f_1,f_2,\ldots, f_N\}$, and try to identify which of these models is best capable of meeting the design objectives, $T$. Here we adopt an inferential perspective, but instead of calibrating a model against observed data we calibrate the models against the data we would like to see, that is, the design objective. 
\par
For each model $f_i\in \mathcal{F}$ the aim is to obtain an estimate for the probability that $f_i$ is the best model capable of producing the target behaviour, given $T$ and the set of candidate models, $\mathcal{F}$,  $\pr(f_i|T)$. This {\em posterior model probability} is given by
\begin{equation}
    \pr(f_i|T) = \frac{\pr(T|f_i)\pi(f_i)}{\pr(T)}= \frac{\pr(T|f_i)\pi(f_i)}{\sum_i \pr(T|f_i)\pi(f_i)}, \label{eq:modelpost}
\end{equation}
where $\pi(f_i)$ is the {\em prior probability} of model $f_i$, $\pr(f_i)$, denotes the {\em likelihood}, and $\pr(T)=\sum_i \pr(T|f_i)\pi(f_i)$, the {\em evidence}, which can be interpreted as a normalising factor (akin to the partition function in statistical physics). 
\par
The terms in Equation \eqref{eq:modelpost} refer to the models. We have subsumed the parameter dependence implicitly into these terms; but we need to consider them, of course, as a model $f_i$ will typically only exhibit the target behaviour, $T$, for certain regions of the parameter spaces, $\Omega_{\theta_i}$. We let the $i$ subscript refer to the model, $f_i$, and we allow different parameter spaces for the different models. The {\em marginal likelihood}, $\pr(T|f_i)$, is given by
\begin{equation}
    \pr(T|f_i) = \int_{\Omega_i} d\theta' \pr(T|f_i,\theta')\pi_i(\theta'), 
\label{eq:parpost}
\end{equation}
where $\pr(T|f_i,\theta)$ is the likelihood of the parameter, and $\pi_i(\theta)$ denotes the prior of the parameters for model $i$ (accounting for the fact that parameter vectors can differ between models).
\par
The rationale for Bayesian design is that it allows us to estimate the probability that a given model, $f_i$ will be able to meet the design objectives, $T$. This is dependent not only on $T$, but also on the set of alternative models considered. The model posterior probability, $\pr(f_i|T)$, can thus be used to choose from the available alternative designs. 
\par
This approach has an intrinsic robustness criterion embedded at its heart: in Equation \eqref{eq:parpost} the integral is taken over the parameter space, which means that  the model, $f_i$, with the maximum {\em a posteriori} probability will generally show good performance against the design objectives over a range of parameters. With this information it is possible to assess the likely robustness -- a concept that we will revisit below -- of a design, and compare competing designs in terms of expected performance and robustness.
\par
\subsection{Approximate Bayesian Computation for Design}
This likelihood is also difficult to evaluate for most important problems of scientific and technological interest. As a result, a range of approximations have been developed that forgo the explicit evaluation of the likelihood in favour of using simulations. Of these, approximate Bayesian computation \cite{Toni:2009tr,Wilkinson:2013ci}, or ABC, is perhaps the best known alternative to evaluating the full posterior via the likelihood function. It has been used in the context of design with some success\cite{Barnes:2011hh} and we outline it below before considering applications, and generalisations. 
\par
In brief, we sample a parameter value, $\theta'$ from the prior, $\pi_i(\theta)$, and we then simulate the model for that parameter value to obtain simulated data $\mathcal{D}=\{d_1,d_2,\ldots,d_n\}$. This is then compared to the real data -- or in the present context to the design objectives, perhaps via surrogate data $\mathcal{D}_T$ -- via a distance function, $\Delta(X,Y)$, such that $\theta'$ is regarded as a draw from the {\em approximate posterior}, if
\begin{equation}
    \Delta(\mathcal{D},T) \le \epsilon
\end{equation}
where $\epsilon$ has to be sufficiently small. The full posterior for the parameter of a given model is thus replaced by
\begin{equation}
    \pr(\theta|T) \approx \pr_{\text{ABC}}(\theta| \Delta(\mathcal{D},T) \le \epsilon),
\end{equation}
and analogously for the model posterior. (Here we can think of sampling models and then model parameters simultaneously). Note that here we do not, as is frequently the case\cite{Robert:2011dk}, take a summary statistic of real and simulated data, but compare the simulation output directly to the data (or the design objectives).
\par
In ABC inference the choice of the distance function is of pivotal importance and can make profound differences on the quality and indeed validity of the inference. Classical rejection-based ABC (where only a single threshold is chosen) tends to be computationally inefficient and wasteful. Instead most inference approaches in the literature adopt a sequential Monte Carlo (SMC) approach, where a sequence of thresholds, $\epsilon_1\ge \epsilon_2\ge\ldots\ge \epsilon_n$ is used (either predefined or constructed ``online" over the course of the inference procedure -- thus either $n$ or $\epsilon_n$ can be a random variable). Choice of this threshold schedule and other aspects of the ABC-SMC approach are still requiring a lot of attention.
\par
In the context of design the distance function also needs to do a lot of heavy lifting, both in terms of performance and, more fundamentally, for the conceptual set-up of the ABC design approach.
\par
In the simplest case we could take an indicator function,
\begin{equation}
    \Delta(\mathcal{D},T) = \begin{cases}0 \qquad \text{ iff $\mathcal{D}$ has property $T$}\\
    0\qquad  \text{ otherwise.}
    \end{cases}
\end{equation}
But if we want to put ABC design to work in the context of the vastly more efficient SMC samplers we have to rethink such indicators. This may not always be possible, of course. The apparent simplicity of ABC approaches masks that there is a great deal of care required to make sure that model selection can be performed in this framework (discussed e.g. in \cite{Robert:2011dk,Barnes:2012hf,Prangle:2014ee} 
\afterpage{
\begin{figure*}
    \centering
    \includegraphics[width=\textwidth]{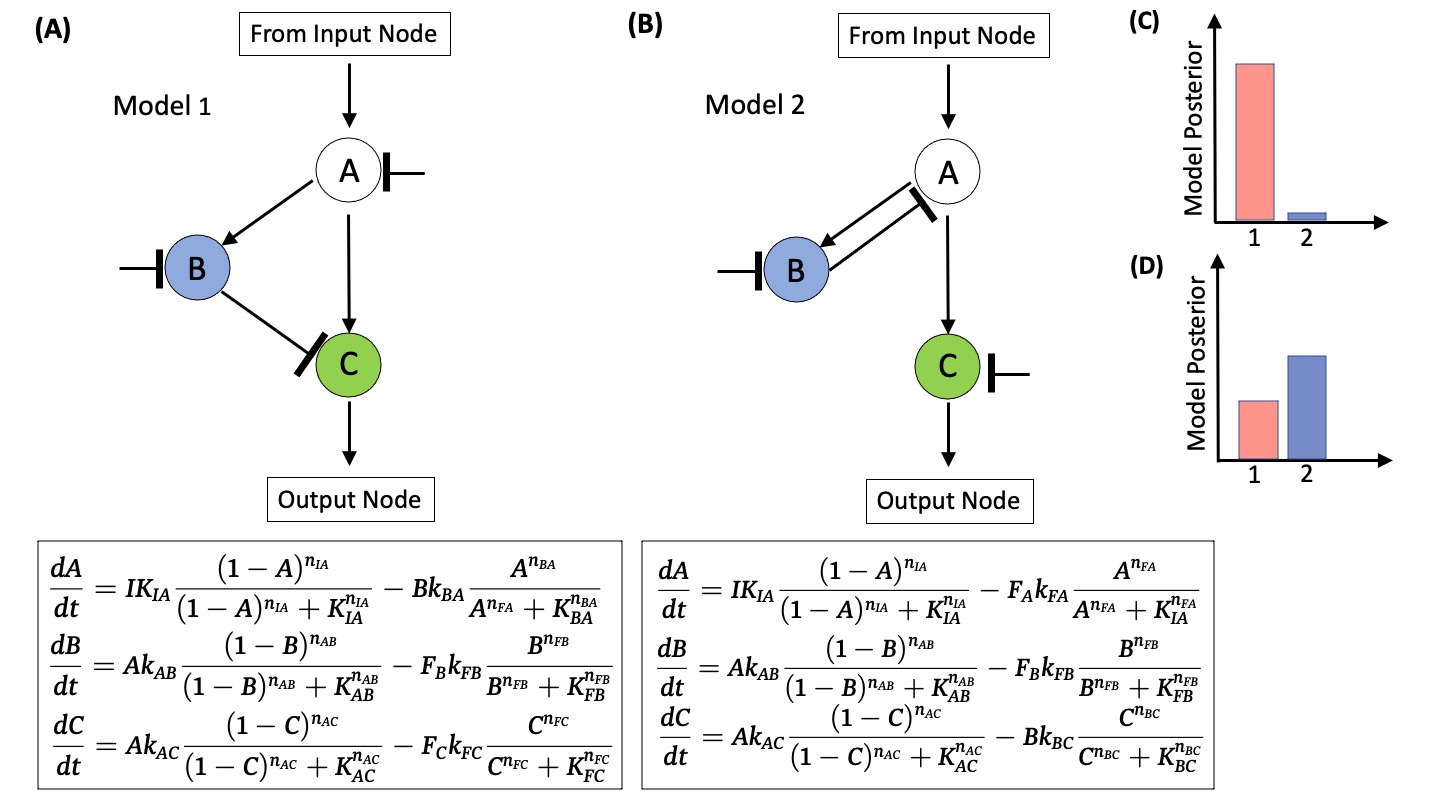}
    \caption{Bayesian design of an adaptive system. (A) and (B) Two models capable of exhibiting adaptive behaviour, including their representations as ordinary differential equations.(C) Posterior model probabilities for achieving adaptive behaviour by models 1 and 2 if there is cooperativity. (D) Posterior model probabilities for achieving adaptive behaviour by models 1 and 2 when there is no cooperativity. The posterior probabilities in (C) and (D) sum up to one.}
    \label{fig:fig_abc}
\end{figure*}
}

\subsection{ABC Design for Perfect Adaptive Behaviour}
We can illustrate the ABC for design approach in the context of adaptive behaviour. In addition to the algebraic methods \cite{Araujo:2018ce}, which
can be used to distill the design characteristics, large-scale computational searches \cite{Ma:2009p20838} have been used. These identified two models that are able to generate adaptive behaviour. Either the algebraic approach or brute-force computational surveys can be used to produce sets of model designs, $\mathcal{F}=\{f_1,\ldots,f_n\}$ which can then be used as the starting point for model selection for design.
\par
The output characteristics in \cite{Barnes:2011hh} were chosen to be the adaptation efficiency and sensitivity \cite{Ma:2009p20838}, given by
\begin{align}
    E=& \frac{(O_2-O_1)/O_1}{(I_2-I_1)/I_1}\\
    \intertext{and}
    S& \frac{(O_{\text{peak}}-O_1)/O_1}{(I_2-I_1)/I_1},
\end{align}
where $I_1$ and $I_2$ are the baseline and activated input levels; $O_2$ and $O_1$ are the steady state outputs before and after the signal is applied; and $O_{\text{peak}}$ is the maximal transient output level.
\par
In Fig. \ref{fig:fig_abc} we show two models (A), (B), that have been argued to exhibit adaptive behaviour in previous studies \cite{Barnes:2011hh,Ma:2009p20838}, and their respective posterior probabilities to produce adaptive behaviour if cooperativity is allowed (C) or not allowed (D). The two posterior probabilities are sufficiently different in the cooperative scenario (C) to warrant choice of model 1 $\pr(f_1|T)\approx 0.96$; whereas for non-cooperative dynamics the choice is less clear with  $\pr(f_2|T)\approx 0.64$. This is not decisive enough and further analysis, and consideration of further design candidates, may become necessary.
\par

\section{Robustness in Synthetic Biology Design}
Below we develop two notions of robustness in synthetic biology design:\\
{\it Biological Robustness} refers to the ability of a system to perform its function in the presence of perturbations.\\
{\it Model Robustness} is related to mismatches between the mathematical model and the real biological systems it sets out to describe.\\
Both are relevant in the context of synthetic and engineering biology and design approaches, and are discussed below.

\subsection{Robustness of Biological Systems and Models}
Robustness as a property of both biological systems and their representations as mathematical models has been the subject of much discussion \cite{Barkai1997,Alon1999,Crampin1999,Little1999,Ma2002,Morohashi2002,Kitano2004a,Tian2004,Rand2008,Raerinne2013,Li2014,Matsuoka2016}. For biological systems, robustness is generally defined as the property that ensures the system maintains its functions in the presence of perturbations from internal and external sources \cite{Kitano2004,Stelling2004,Kitano2007}. In the context of design we take a pragmatic point of view of robustness. We are interested in identifying designs that are robust to (i) inherent stochasticity; (ii) environmental and physiological signals and fluctuations; (iii) uncertainty in our knowledge of the precise system architecture and kinetic reaction rates.
\par
For mathematical models many working definitions of robustness have been proposed, which are based on single- and multi-parameter sensitivity analysis \cite{Ma2002,Matsuoka2016}. They include, for example, invariance under external fluctuations affecting the initial conditions \cite{Crampin1999}; or structural stability \cite{Blanchini2011}. There is, however, no clear and consistent definition that allows for universal and consistent quantification of robustness, which is necessary if the  robustness characteristics of different systems are to be compared. Perhaps the primary importance of quantifying model robustness is in model validation, as a valid model must exhibit equivalent robustness to the corresponding biological system it represents. In fact, a lack of robustness may indicate a weak model that provides a poor representation of the biological system \cite{Morohashi2002}, so robustness may be used as a selection tool for promising models that correctly represent the functions of the biological system. The determination of weak models through robustness quantification is also useful for providing insight into the robustness of the biological system in a synthetic biology design context. 
\par
Importantly, robustness relates to the maintenance of the functions of the system and not necessarily the states of the system, so is more general than stability and homeostasis and may involve transitions to new steady states or even to an instability if required to maintain function \cite{Kitano2007}. For example, system function where a morphogen concentration gradient determines the position in a biological system appears to be robust despite temporal perturbations of the gradient to different steady-states \cite{Tostevin2007}. Despite their complexity, evolved biological systems apparantly exhibit robustness as a ubiquitous and fundamental property \cite{Kitano2004,Wagner2008,Young2017}. Synthetic cellular systems have to work in the context of such naturally evolved contexts.
\par

\subsection{Mathematical Definitions of Robustness}
Unfortunately, and as noted some time ago \cite{Kitano2007}, there is still no adequate mathematical theory of robustness. There are a number of analytical issues to be addressed, in particular the difficulty involved with high-dimensional parameter spaces. For example, control-theoretic techniques have been employed to study the robustness of biochemical network models \cite{Ma2002}; however, these approaches are limited to low-dimensional parameter spaces \cite{Kitano2004}. 
\par
A robust model must maintain the represented biological function with respect to some degree of perturbation of the inputs. In other words, we require small changes in the inputs -- model parameters, intitial conditions, boundary conditions, external conditions -- to produce sufficiently small changes in the output so that the integrity of the biological function remains. Note that this is different from requiring invariance of the output: the  system output can vary as long as the biological function is maintained. We therefore only require invariance of biological function: the design objectives, $T$ still have to be met. 
\par
We require a way of assessing robustness and focus here, for didactic reasons, on robustness to parametric uncertainty or variation. In order to compare distances between input values and output values (including any design objective, $T$) we require the sets of input and output values to form metric spaces.
\par
We furthermore require a functional relationship between the input and output metric spaces, which would generally be a continuous function in order to ensure that the output values change smoothly with changes in the inputs. {It is also possible that an appropriate function has discontinuities, such as a step discontinuity, so may be only piecewise continuous. In this case we just need to ensure that the function output still corresponds to invariance of biological function by  accounting for the behaviour near the discontinuities. This makes it possible to accommodate qualitative design criteria.} We would also like to be able to specify the robustness of a system by a single parameter; we therefore need a methodology that can be applied to the whole parameter space to give an overall measure of robustness.
\par
From a global perspective, we require that close values of the input produce output values that are consistently close. Note that local continuity of a function, or global uniform continuity, are not sufficient for defining robustness as they allow for arbitrarily large derivatives of the function, and thus a lack of robustness. The Lipschitz function \cite[Section 9.4, page 154]{OSearcoid2007} immediately lends itself to a rigorous definition of robustness for mathematical models. But inexplicably it has not been employed within the mathematics literature, to our knowledge; it has, however, appeared in other fields such as software engineering \cite{Chaudhuri2012} and cyber-physical systems \cite{Majumdar2011}. 
\par
We denote the input and outputs by $X$ and $Y$, respectively; and we define suitable metrics, $d_X$ and $d_Y$ on the statespaces, for $X$ and $Y$; we can therefore consider $(X,d_X)$ and $(Y,d_Y)$ to be metric spaces. A function $f \colon X \to Y$ is a \emph{Lipschitz function} on $X$ with a \emph{Lipschitz constant} $K \ge 0$ if $f$ satisfies
\begin{equation}
d_Y (f(x_1),f(x_2)) \le K \, d_X (x_1,x_2)
\label{eq:Lip}
\end{equation}
for all $x_1$, $x_2 \in X$. We also say that $f$ is \emph{Lipschitz continuous}. The smallest such constant $K$ is \emph{the} Lipschitz constant. The Lipschitz constant provides an upper bound on the change of the function output given a change in the input. Ideally we would like an analytical value for the Lipschitz constant, however there may be Lipschitz functions for which this is not easily obtained, so numerical estimates may be helpful. Numerical methods for estimating Lipschitz constants exist \cite{Wood1996}; however, there is no guarantee that the estimated constant is valid globally. In control theory \emph{transfer functions} are employed  to provide a functional relationship between the inputs and outputs of a system, often within the frequency domain, by applying the Laplace transform to both the inputs and outputs, thereby allowing direct comparison of functions \cite{Zabczyk2020}. While the overall approach using transfer functions overlaps with our requirements, it is perhaps too specific and limited for characterising the robustness of complex biological systems.
\par

\subsection{Parametric Robustness as a Design Criterium in Synthetic Biology}
The algebraic approaches outlined above allow us to detect the design principles underlying a given desired behaviour or objective, $T$. This allows us to design models that meet the objectives by construction. 
\par
The Bayesian approach contains an assessment of robustness via the posterior\cite{Secrier:2009ko}, which integrates over the parameters, Equation \ref{eq:parpost}. Rather than the design objective, $T$, we may use the posterior parameter density, $\pr(\theta|T)$, or the ABC posterior, \eqref{eq:parpost}, as the output, and then compare $d_\theta(\theta,\theta')$ and $d_P(\pr(\theta|T),\pr(\theta'|T)$ using the Lipschitz criterion \eqref{eq:Lip}. We can focus this analysis on, for example, the 90\% credible interval \cite{Robert:2007uy}.
\par
If our design, that is the mathematical model, is robust against variation in parameters, if it is structurally stable, and if is is even robust against changes to the model structure \cite{Babtie:2014jg}, then we can view this as reassuring that the design will also be robust to the discrepancies between mathematical model and biological, reality\cite{Kirk:2015gj,Scholes:2019fq}. 

\section{Discussion and Conclusions}
The models in section III and in Figure \ref{distributedintegrals} are structurally very similar to the models considered in section V and Figure \ref{fig:fig_abc}, but differ in practical terms as we first use mass action kinetics and then Michaelis-Menten kinetics. Algebraic methods are more powerful wherever they are applicable, as they result in definitive criteria that must be satisfied for a given behaviour to emerge. From this we can often gain profound insights into the dynamics of cellular systems \cite{Araujo:2018ce,Harrington:2013wv}. 
\par
For many classes of design objective, other methods need to be applied, and the Bayesian design framework is one method that is widely applicable for different models. Basing design on  Bayesian model selection \cite{Toni:2010p29729,Barnes:2011bka,Barnes:2011hh} has two intrinsic advantages: (i) we are able to balance the performance and the robustness of different designs; and (ii) for ABC methods we can, at least in principle, apply Bayesian design to arbitrary models as long as we are able to simulate them. The first point is potentially an advantage over large-scale or exhaustive analyses of models \cite{Ma:2009p20838,Schaerli:2014hr,Babtie:2014jg,Scholes:2019fq}. There, robustness can be assessed {\em post-hoc}. While global analyses and Bayesian methods are both computationally expensive, the former are better suited for the comprehensive analysis of a large number of systems: the objective there is to scan model and parameter spaces to find if there are any parameters for which the design objectives, $T$, are realised. In the Bayesian approach, by contrast, we also obtain estimates of the parameters, $\theta_i$, for each model $f_i\in\mathcal{F}$.   
\par
So we can view the algebraic and Bayesian frameworks as complementary, and given the complexities of rationally designing synthetic circuits \cite{Lim:2010p24929,Purcell:2013iq} a flexible and diverse tool-box of alternative design methods is clearly beneficial.  But algebraic methods can also support and work in conjunction with Bayesian methods (or exhaustive scans of large sets of models): we can determine initial sets of models, $\mathcal{F}=\{f_1,\ldots,f_n\}$ to fulfil the design criteria based on algebraic rules or design principles (even if under perhaps restricted kinetic regimes, such as mass-action kinetics); a natural choice for the prior model probabilities would then be, $\pi(f_i) = 1/n$ for all $i=1,\ldots,n$. These models can then be subjected to the Bayesian model selection/design process to choose those models that have the highest probability of delivering the design objective, $T$ (and do so robustly). 
\par
The challenge of design in synthetic biology is considerable. Unlike in physics and many engineering disciplines we cannot draw on first-principles. Reliable mathematical models for cellular systems are still, despite much recent progress, rare, but we need them to apply rational engineering design approaches in synthetic biology. The methods described here -- algebraic ways to identify design principles; Bayesian design; and the concept of robustness -- can help to identify synthetic circuits capable of delivering  a target objective, $T$. In combination they allow us to take some of the guess-work and trial-and-error out of synthetic biology design.

\ifCLASSOPTIONcaptionsoff
  \newpage
\fi



\bibliographystyle{IEEEtran}
\bibliography{./bibtex/bib/IEEEexample.bib}
\end{document}